\documentclass[aps,prl,twocolumn,showpacs,superscriptaddress,groupedaddress, floatfix]{revtex4} 
\usepackage{amsmath}
\usepackage{amssymb}
\usepackage{amstext}
\usepackage{amsopn}
\usepackage{amsfonts}
\usepackage{amsxtra}
\usepackage[english]{babel}
\usepackage{graphicx}
\usepackage{bm}
\usepackage{multirow}
\usepackage{dcolumn}
\usepackage{color}
\usepackage{todonotes}
\usepackage{verbatim}
\usepackage{textcomp}
\usepackage{soul}

\begin{document}

\title{Provoking topology by octahedral tilting in strained SrNbO$_3$}

\author{A.~Chikina}
\thanks{chikina.alla@gmail.com}
\address{Photon Science Division, Paul Scherrer Institut, CH-5232 Villigen, Switzerland}
 \author{V.~Rosendal}
 \address{Department of Energy Conversion and Storage, Technical University of Denmark, 2800 Kgs. Lyngby, Denmark}
  \author{H.~Li}
 \address{Photon Science Division, Paul Scherrer Institut, CH-5232 Villigen, Switzerland}
\author{E.~B.~Guedes}
 \address{Photon Science Division, Paul Scherrer Institut, CH-5232 Villigen, Switzerland}
 \author{M.~Caputo}
 \address{MAX IV Laboratory, Lund University, P.O. Box 118, 22100, Lund, Sweden}
 \author{N.~C.~Plumb}
 \address{Photon Science Division, Paul Scherrer Institut, CH-5232 Villigen, Switzerland}
 \author{M.~Shi}
 \address{Photon Science Division, Paul Scherrer Institut, CH-5232 Villigen, Switzerland}
 \author{D.~H.~Petersen}
 \address{Department of Energy Conversion and Storage, Technical University of Denmark, 2800 Kgs. Lyngby, Denmark}
 \author{M.~Brandbyge}
 \address{Department of Physics, Technical University of Denmark, 2800 Kgs. Lyngby, Denmark}
 \author{W.~H.~Brito}
 \address{Departamento de Física, Universidade Federal de Minas Gerais, C. P. 702, 30123-970, Belo Horizonte, MG, Brazil}
 \author{E.~Pomjakushina}
 \address{Laboratory for Multiscale Materials Experiments, Paul Scherrer Institut, CH-5232 Villigen, Switzerland}
 \author{V.~Scagnoli}
 \address{Laboratory for Multiscale Materials Experiments, Paul Scherrer Institut, CH-5232 Villigen, Switzerland}
  \address{Department of Materials, Laboratory for Mesoscopic Systems, ETH Zurich, 8093 Zurich, Switzerland}
 \author{J.~Lyu}
 \address{Laboratory for Multiscale Materials Experiments, Paul Scherrer Institut, CH-5232 Villigen, Switzerland}
 \author{M.~Medarde}
 \address{Laboratory for Multiscale Materials Experiments, Paul Scherrer Institut, CH-5232 Villigen, Switzerland}
 \author{E.~Skoropata}
 \address{Photon Science Division, Paul Scherrer Institut, CH-5232 Villigen, Switzerland}
 \author{U.~Staub}
 \address{Photon Science Division, Paul Scherrer Institut, CH-5232 Villigen, Switzerland}
 \author{S.-W.~Huang}
 \address{Photon Science Division, Paul Scherrer Institut, CH-5232 Villigen, Switzerland}
 \author{F.~Baumberger}
 \address{Photon Science Division, Paul Scherrer Institut, CH-5232 Villigen, Switzerland}
  \address{Department of Quantum Matter Physics, University of Geneva, 24 Quai Ernest-Ansermet,1211 Geneva 4, Switzerland}
 \author{N.~Pryds}
 \thanks{nipr@dtu.dk}
 \address{Department of Energy Conversion and Storage, Technical University of Denmark, 2800 Kgs. Lyngby, Denmark}
 \author{M.~Radovic}
 \thanks{milan.radovic@psi.ch}
 \address{Photon Science Division, Paul Scherrer Institut, CH-5232 Villigen, Switzerland}

\date{\today}
\title{Provoking topology by octahedral tilting in strained SrNbO$_3$}

\begin{abstract}
Transition metal oxides with a wide variety of electronic and magnetic properties offer an extraordinary possibility to be a platform for developing future electronics based on unconventional quantum phenomena, for instance, the topology. The formation of topologically non-trivial states is related to crystalline symmetry, spin-orbit coupling, and magnetic ordering. Here, we demonstrate how lattice distortions and octahedral rotation in SrNbO$_3$ films induce the band topology. By employing angle-resolved photoemission spectroscopy (ARPES) and density functional theory (DFT) calculations, we verify the presence of in-phase $a^0a^0c^+$ octahedral rotation in ultra-thin SrNbO$_3$ films, which causes the formation of topologically-protected Dirac band crossings. Our study illustrates that octahedral engineering can be effectively exploited for implanting and controlling quantum topological phases in transition metal oxides. 
\end{abstract}

\maketitle


\section{Introduction}
The physical properties of perovskite-type ABO$_3$ transition metal oxides (TMOs) are largely determined by BO$_6$ octahedra that form a three-dimensional network. Variations in the size, symmetry, and orientation of octahedra open the possibility of creating new properties due to the strong coupling between the lattice, charge, spin, and orbital degrees of freedom in TMOs~\cite{Rondinelli2010, Rondinelli2011,Liu2013,Moon2014,Kan2016,Liao2016,Zhaoliang2018}.
In recent years, there has been a growing interest in exploring topological phases in TMOs, which requires to the breaking of certain symmetries, such as inversion or time-reversal symmetries~\cite{Hirschmann2021,Rondinelli2010, Mohanta2021}. In this regard, TMO thin films possess great potential for creating novel quantum phases by tailoring structural distortions~\cite{Rondinelli2012}. 
For instance, it was predicted that cubic perovskite oxides could be transformed into a tetragonal structure with non-symmorphic symmetry in epitaxially-strained thin films~\cite{Rondinelli2010, Mohanta2021, Herklotz2016}. This induced octahedral rotations, which lower the symmetry, cause a band folding and may give rise to a Dirac crossing near the Fermi level. These massless fermions are protected~\cite {Hirschmann2021}, leading to higher carrier mobility, which is a key property for high-speed electronics~\cite{Liang2015, Schoop2016, Arnold2016, Yang2020}.

\begin{figure}
\centering
\includegraphics[scale=0.4]{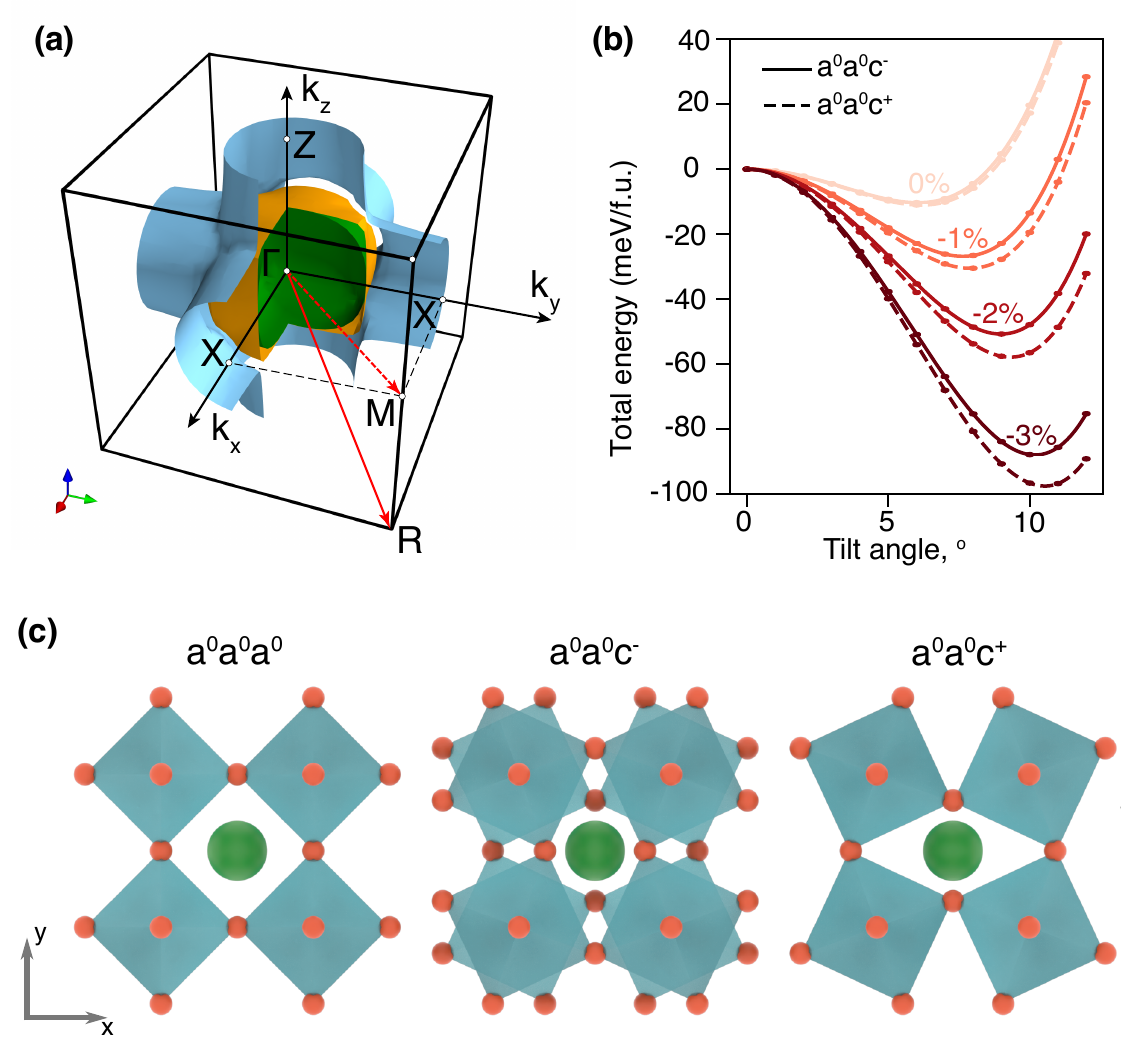}
 \caption{({a}) Rendering of the DFT 3D Fermi surface of the cubic phase determined by DFT calculations. Red arrows are reciprocal lattice vectors of the supercell arising from the $a^0a^0a^+$(dashed line) and $a^0a^0a^-$(solid line) tilting modes. They illustrate the band folding for both configurations. 
 ({b}) Strain-tilt energy landscape. The strain of -2 $\%$ is applied in the $xy$-plane. The total energy difference is relative to the $a^0a^0a^0$ structure and is normalized to the unit cell. ({c}) Sketches of rotation modes with exaggerated angles.} 
 \label{fig:Fig1}
\end{figure}

Band structure analysis of tetragonal perovskite oxides~\cite{Mohanta2021}, including CaNbO$_3$, SrRuO$_3$, SrMoO$_3$, SrTiO$_3$, and SrNbO$_3$, suggests that the non-trivial Dirac crossing is located closest to the Fermi level in SrNbO$_3$ with out-of-phase $a^0a^0c^-$ octahedral rotation (using the Glazer notation~\cite{Glazer1972}, see Fig.\ref{fig:Fig1}c). 
Another theoretical study predicts that in-phase $a^-a^-c^+$ octahedral rotation also induces a topological band structure in  SrNbO$_3$~\cite{Zhang2019}. While both in-phase and out-of-phase octahedral rotations lead to the formation of symmetry-protected Dirac crossings, the position in energy and reciprocal space are different due to the specific band folding. 
A recent study of SrNbO$_3$ thin films found out-of-phase $a^0a^0c^-$ octahedral rotation and reported evidence for a non-zero Berry curvature from an analysis of quantum oscillations~\cite{ok2021}. Generally conveying, understanding how the structural change in TMOs affects the band structure is of utmost importance to utilize topological phases in this material. The most direct method to reveal the band structure topology in relation to crystal symmetry is angle-resolved photoemission spectroscopy (ARPES). In our study, we combine ARPES and density functional theory (DFT) calculations to investigate the influence of octahedral rotations on the band topology and the evolution of the electronic structure in SrNbO$_3$ films. 

\section{Results and discussion}
SrNbO$_3$ is a transparent conductor in the visible and ultraviolet range~\cite{Park2020, Daichi2015}. The stoichiometric material has one electron in the Nb 4$d$ states distributed over the three almost degenerate t$_{2g}$ orbitals. Each orbital forms a cylindrical Fermi surface sheet along a principal axis, and weak hybridization results in the three Fermi surfaces (illustrated in Fig.\ref{fig:Fig1}a for the cubic phase $a^0a^0c^0$ SrNbO$_3$.)
In our previous work, we analyzed possible, stable configurations of the strained SrNbO$_3$ by performing first-principles DFT calculations ~\cite{rosendal2023}. Among the various studied octahedral configurations~\cite{rosendal2023}, this study is focused on $a^0a^0c^+$ and $a^0a^0c^-$, which are the most stabilized under compressive biaxial strain (Fig. \ref{fig:Fig1}b,c).  
Interestingly, DFT calculations show that both studied octahedral configurations of SrNbO$_3$ are near-degenerate with the cubic one while getting more stable with increasing the strain (Fig. \ref{fig:Fig1}b). The energetic stabilization of the distorted structures is further favorable by the moderate electronic correlations raised in SrNbO$_3$, as pointed out in our previous study~\cite{rosendal2023} based on DFT$+$DMFT calculations.

\begin{figure}
\centering
\includegraphics[width=0.5\textwidth]{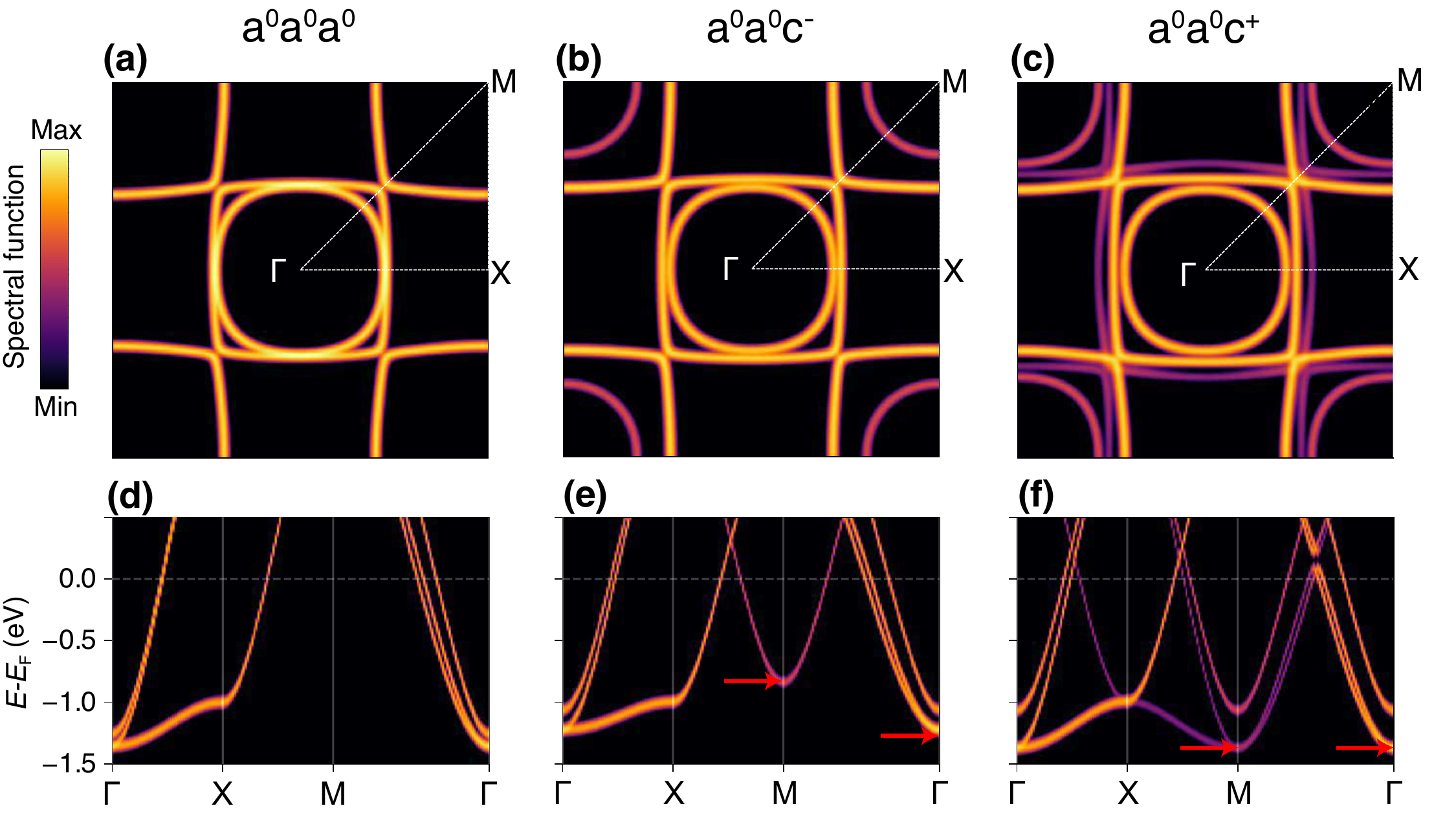}
\caption{Evolution of the DFT Fermi surface slices and band structures of SrNbO$_3$ with octahedral rotation for ({a,d}) $a^0a^0a^0$, ({b,e}) $a^0a^0c^-$ and ({c,f}) $a^0a^0c^+$ phases. Calculations are done for a -2 $\%$ strained films with a 9~$ ^{\circ}$ rotation. Panels ({a-b}) present unfolded Fermi surface slices at $k_z = 0$, and panels ({d-f}) show unfolded band structures of the different tilt modes under -2\% in-plane biaxial strain. Note that we use the cubic  BZ notation and a logarithmic color map. Note that the spectral function here is a measure of the Bloch character of the states smeared using a Gaussian~\cite{Popescu2010,Popescu2012}. Furthermore, the smearing is arbitrary and does not reflect any finite lifetimes due to many-body effects.}
 \label{fig:Fig3}
\end{figure}

The calculated band structures for the pseudocubic $a^0a^0c^0$ and two out-of-phase $a^0a^0c^-$ and in-phase $a^0a^0c^+$ modes are presented in Fig.~\ref{fig:Fig3}. To simplify the comparison, we unfold the phases onto the cubic $a^0a^0a^0$ lattice and use this notation of the Brillouin zone (BZ). 
Below, we investigate the distinction in the band structure of both phases, using the cubic notation of BZ. 


In order to capture the band structure of $a^0a^0c^-$, the $a,b$, and $c$ lattice parameters of the cubic need to be multiplied by a factor of $\sqrt{2}$. This new periodicity causes band folding (translation vector indicated by the solid arrow in Fig.\ref{fig:Fig1}a). Consequently, the new pocket at the ${M}$-point is formed by folding the pocket at the ${Z(X)}$-point. In addition, due to a mirror folding of the band structure in the cubic BZ inside the smaller tetragonal BZ, a Dirac-like crossing is formed along ${X}$ - ${M}$ and $\Gamma$ - ${M}$ directions. More details regarding band foldings are presented in the Supplementary Information.

\begin{figure}
\centering
 \includegraphics[scale=0.12]{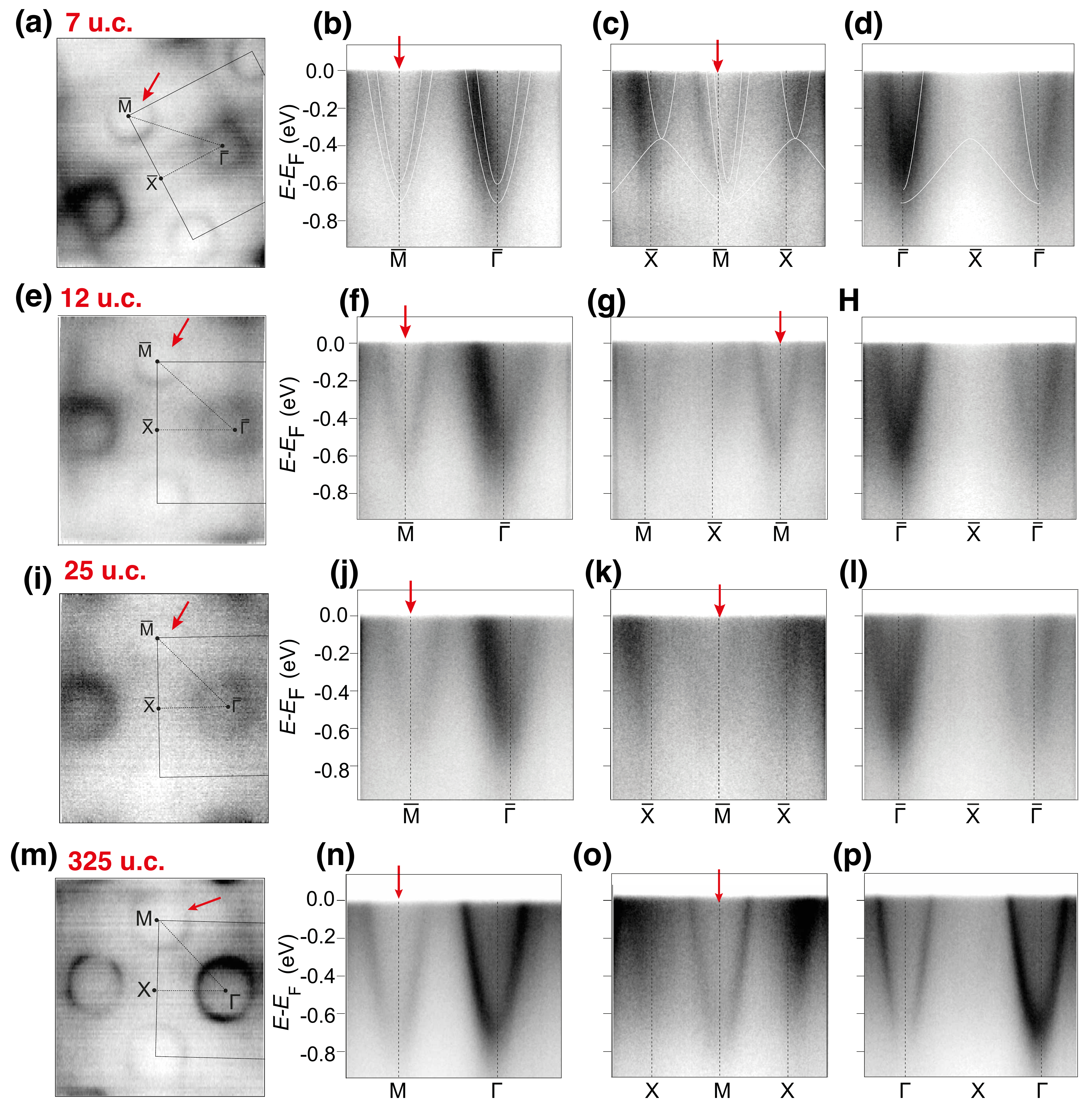}
 \caption{Spectra taken at 106 eV photon energy and circular polarisation of {(a-d)} 7 u.c., {(e-h)} 12 u.c., {(i-l)} 25 u.c.
 and {(m-o)} 325 u.c. thick SrNbO$_3$ films. DFT band structure shifted toward Fermi level by 0.6 eV is presented by thin white lines in panels ({b-d}). Surface-projected symmetry points are marked with a dashed line. The red arrow marks an additional pocket around ${M}$-point, which appears due to the octahedral rotations.}
 \label{fig:Fig2}
\end{figure}

In the case of  $a^0a^0c^+$ in-phase octahedral rotation, the cubic cell needs to multiply by a factor $\sqrt2\times\sqrt2\times1$ (see Supplementary information). That results in reducing and 45$^o$ rotation of the BZ  (P4/mbm (127) space group) and the folding of the "cubic" band structure over  the ${X}-{X}$ diagonal plane (dashed red arrow in fig.\ref{fig:Fig1}a). This folding results also in  Dirac-like band crossings along $\Gamma$ - ${M}$,  ${X}$ - ${M}$ and $\Gamma$ - ${X}$ directions (Fig~\ref{fig:Fig3}c,f) and translation of the band structure from the $\Gamma$-point to the ${M}$-point (dashed arrow in Fig.\ref{fig:Fig2}a). Importantly, the new pocket at ${M}$-point  either corresponds to the original band from  ${\Gamma}$- or ${X}$- points, depending on the octahedral rotation phase. According to the calculations, the folded band at the ${M}$-point is located at lower energy for the in-phase $a^0a^0c^+$ (indicated by arrows in Fig~\ref{fig:Fig3}e,f). Therefore, the position of the band bottom at ${M}$-point and the presence  of the heavy band along ${X}$ - ${M}$ direction are fingerprints of the octahedral configuration type, allowing us to distinguish them. 

To identify which phase is stabilized, we measured the band structure of SrNbO$_3$ films using ARPES (Fig.~\ref{fig:Fig2}).
All measured films were grown by pulsed laser deposition at the Swiss Light Source's Surface/Interface Spectroscopy (SIS) beamline at the Paul Scherrer Institut in Villigen, Switzerland. The films were grown on TiO$_2$-terminated SrTiO$_3$ (001) substrates at T = 650°C and P(O$_2$) $<$ 5 × 10$^{-7}$ mbar, resulting in single-phase epitaxial films (see Supplementary Information). The lattice mismatch between the SrNbO$_3$ film and SrTiO$_3$ (001) substrate induces a compressive strain of about 2 \% in the film and presumably octahedral rotations\cite{ok2021}. After the growth, the films were transferred \textit{in situ} to the  ARPES station at the SIS beamline via an ultra-high vacuum transfer line.

Previous photoemission studies of the electronic structure of 15 nm (around 38 unit cells) SrNbO$_3$ films grown on DyScO$_3$ (110) showed no indications of deviation from the cubic SrNbO$_3$ structure~\cite{Bigi2020}.
Nevertheless, SrNbO$_3$ film grown on DyScO$_3$ (110) should experience compressive strain, but it can relax after several tens of unit cells \cite{Truttmann2021, Siwakoti2021}. Therefore, to observe the conceivable strain effect, we studied the band structure evolution for SrNbO$_3$ films of different thicknesses: 7, 12, 25, and 325 unit cells (u.c.). 
In the following part, we focus on two SrNbO$_3$ films: ultra-thin 7 u.c. film, where the strain should have a more significant effect, and a thick film of 325 u.c. (130 nm), where the strain effect is minimized through relaxation. Indeed, X-ray reciprocal space mapping confirmed that 25 u.c. film is fully strained (see Supplementary Information). Based on that, we assume that the SNO films with lower thicknesses are also fully strained. Nevertheless, the X-ray reciprocal space mapping shows that 325 u.c. thick (bulk-like) film is fully relaxed (see Supplementary Information). 
Fig.~\ref{fig:Fig2} displays ARPES spectra taken along high-symmetry directions with circularly polarized light of $h\nu=106$ eV photon energy. Fig.~\ref{fig:Fig2}(a,e,i,m) shows the Fermi surface maps in the $\Gamma$-${X}$-${M}$ plane for the films. Besides one spherical and two cylindrical Fermi surface sheets near the $\Gamma$-point, the data reveal a new pocket around the $M$-point for both films. Since the band structure folding occurs for the ultra-thin films and the thick, relaxed film, it is obvious that octahedral rotations are inherent in the SrNbO$_3$ film. However, there is still the question of which type of tilting phase, out-of-phase ($a^0a^0c^-$) or in-phase ($a^0a^0c^+$ ), is present. 





Differences in the band structure between  $a^0a^0c^+$ or $a^0a^0c^-$ configurations allow a more detailed understanding of the crystal structure in SrNbO$_3$ films. The most noticeable distinctions (in Fig.\ref{fig:Fig3} between panels D and F) are the position of the band bottom at the ${M}$  point, and the presence of a weak dispersive folded band along ${X}$ - ${M}$ direction in the $a^0a^0c^+$ configuration. However, we do not observe this band in ARPES spectra, probably due to the matrix element effect. In fact, this band, primarily of $d_{yz}$ character, was noted to be scarcely visible in the ARPES spectra of comparable perovskite transition metal oxides~\cite{Backes2016, Cappelli2022, Takatsu2020, Yoshida2010}.
Therefore, we concentrate on the relative positions of the band minima at the ${\Gamma}$ and ${M}$-points. In our DFT calculations of $a^0a^0c^-$, the conduction band minimum at the ${M}$-point is located $\sim$ 0.35 eV above the ${\Gamma}$-point minimum, whereas $a^0a^0c^+$ phase, the band minimum at ${M}$ and ${\Gamma}$ have the same energy (red arrows in Fig.\ref{fig:Fig3}f). Accordingly, the relative position of band minima at the ${M}$ and ${X}$ points of  $a^0a^0c^-$ and $a^0a^0c^+$ is distinguishable.

In the ARPES spectra of the 7~u.c. film (Fig.~\ref{fig:Fig2}b), which is highly strained (see Supplementary Information), we find that the conduction band minima at ${\Gamma}$ and ${M}$ are around $-0.52\pm0.05$ and $-0.54\pm0.05$ eV, respectively. At the same time, the band minima at ${X}$ is located around $-0.37\pm0.05$ eV (see Fig.~\ref{fig:Fig2}(b-d)). Comparing DFT calculations with experimental results where the band minima at ${\Gamma}$ and ${M}$ points coincide, we deduce that band structure reconstruction originates from $a^0a^0c^+$ in-phase octahedral rotation. This observation is in agreement with our calculation, which predicts that $a^0a^0c^+$ configuration of SrNbO$_3$ under 2\% compressive strain is slightly energetically favorable (see Fig.~\ref{fig:Fig1}b). Nevertheless, we performed synchrotron-based X-ray diffraction measurements to examine the type of phases present in relaxed 325 u.c. thick film. Interestingly, the data analyses of superstructure peaks (See Supplementary Information) reveal the existence of both in-phase and out-of-phase octahedral rotations. Since ARPES is a predominantly surface-sensitive method, we argue that the $a^0a^0c^-$ configuration is not encountered within the surface region.

In Fig.~\ref{fig:Fig2} (b-d), we show the DFT band structure of $a^0a^0c^+$ phase superimposed with experimental spectra of 7u.c.~SrNbO$_3$ film. We find that the calculated bands need to be shifted by 0.6 eV towards the Fermi level to overlap with the experimental data. DFT-based estimation suggests that such a shift corresponds to $\sim 0.58$~$\Bar{e}$ hole doping and thus Nb 4d$^{0.42}$ configuration, instead of Nb 4d$^{1}$ (see Supplementary Information). 
Indeed, X-ray photoemission spectrum of core levels obtained using 650 eV photon energy shows the existence of mixed Nb$^{5+}$ and  Nb$^{4+}$ state (see Supplementary Information). This experimental observation indicates a significant depletion of electrons at the Nb site of SrNbO$_3$~\cite{Bigi2020}. Nevertheless, X-ray diffraction data from 25 u.c. thick films do not show the presence of any other phases (see Supplementary Information). Therefore, the observed hole doping is likely related to self-doping or non-stoichiometry, either from Sr-vacancies or excess oxygen, which is known to exist in SrNbO$_3$ thin film\cite{Bigi2020, Thapa2022}. However, the conduction band minimum in 320 u.c spectra (See Fig. \ref{fig:Fig2}(f,g)) is located at lower binding energy than for the ultra-thin film. As a result, Fermi surface pockets at $\Gamma$ and $M$ point are larger in the case of 325 u.c. film (Fig.\ref{fig:Fig2} m vs a). This can be because the thick film is less hole-doped and probably closer to the stoichiometric composition. Nevertheless, the matching position of the conduction band minima at ${\Gamma}$ and ${M}$ points suggests that $a^0a^0c^+$ phase subsists in the relaxed 130 nm thick film.
\begin{figure}
\centering
\includegraphics[width=0.45\textwidth]{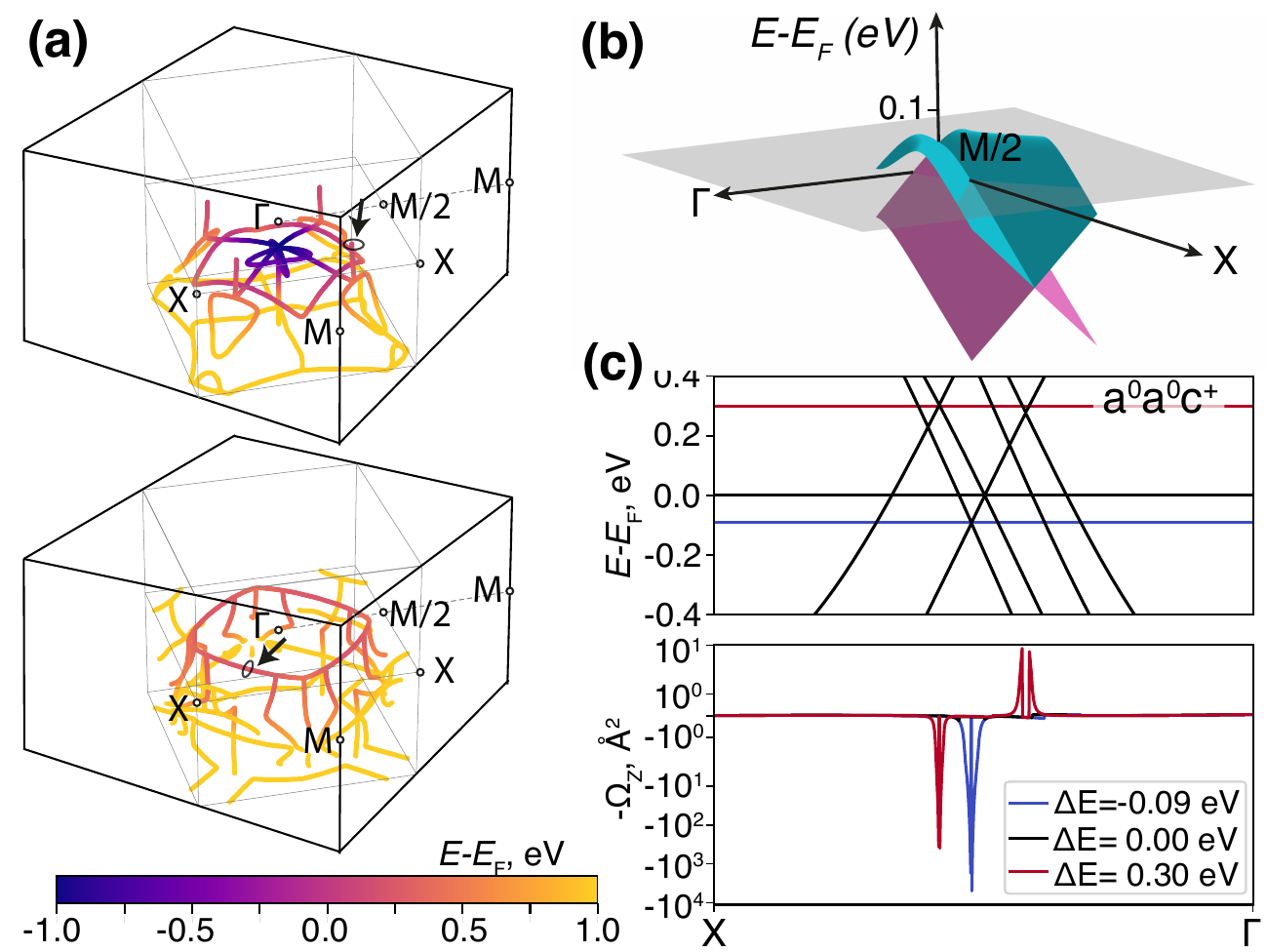}
 \caption{Topology of the SrNbO$_3$ $a^0a^0c^+$ band structure. The BZ is drawn within the BZ of the untilted structure and we use the notation of the untilted structure. (a) Nodal lines (without SOC) of two bands close to Fermi level. A Berry phase of $\pi$ is observed around the drawn circles enclosing the nodal lines. (b) Symmetry protected Dirac nodal line on the boundary of the $a^0a^0c^+$ BZ with SOC included. (c) Bands and Berry curvatures, when time-reversal symmetry is broken by magnetic ordering and SOC, is included. The different lines correspond to different energies relative to the Fermi level.}
 \label{fig:Fig4}
\end{figure}

To investigate the topological properties of the Dirac-like crossings in distorted SrNbO$_3$ ($a^0a^0c^+$ tilt mode), we analyzed the calculated band structure more closely employing the calculation of Berry curvatures and phases, as similarly performed by Mohanta and co-workers~\cite{Mohanta2021}.
Such examination is done in two steps: first, by keeping time-reversal symmetry, and second by breaking the time-reversal symmetry through the introduction of an artificial magnetic ordering (without any external magnetic field).

Within the first step, we examine the properties of the nodal lines in the presence of time-reversal symmetry. Fig.~\ref{fig:Fig4}a displays Dirac nodal lines for two bands, showing energy distribution close to the Fermi level. We considered bands that are populated by the 6 Nb $t_{2g}$ states (2 Nb ions and 3 states per ion). We do not consider other bands with nodal lines since they are further away from the Fermi level. The small circles enclosing the nodal lines with a fourfold degeneracy in Fig.~\ref{fig:Fig4}a show a $\pi$ Berry phase, which indicates a non-trivial topology of these bands~\cite{Fang15}. Furthermore, since the $a^0a^0c^+$ structure belongs to the P4/mbm (127) space group, symmetry-enforced Dirac nodal lines are located along the edge of the BZ~\cite{Yang18}. In this space group, the protected nodal lines are on the ${\mathrm{MX}}$, ${\mathrm{AM}}$ and ${\mathrm{AR}}$ lines~\cite{Hirschmann2021} (in the notation of the P4/mbm (127) space group). An example of such a Dirac nodal line is shown in Fig.~\ref{fig:Fig4}b. Note that the nodal lines protected by the time-reversal symmetry and the nonsymmorphic symmetry are robust even in the presence of SOC\cite{Yang18,Hirschmann2021}.


Next, we break the time-reversal symmetry by creating a magnetic ordering. Fig.~\ref{fig:Fig4}c shows the Wannier interpolated band structure of $a^0a^0c^+$ together with the $z$-component of the Berry curvature. It reaches the most significant values in regions where the folded bands (due to octahedral tilting) are crossing. Here, we constrained a spin polarization in the DFT calculation with a ferromagnetic ordering where each Nb ion has 20\% spin polarization, as conducted in Ref.~\cite{Mohanta2021}. Experimentally, this might be achievable by doping with magnetic impurities or applying an external magnetic field. As expected, the magnetization lifts the spin-degeneracy and breaks time-reversal symmetry, which is necessary for a non-zero Berry curvature in centrosymmetric crystals~\cite{Di10}. The finite Berry curvature can give rise to features such as the anomalous Hall effect~\cite{Naoto10}. 
One of the non-trivial Dirac crossings appears at -0.09 eV below the Fermi level in Fig.~\ref{fig:Fig4}c. It could be used for the realization of novel correlated topological phases. Although our calculations show that the non-trivial Dirac crossing is occupied, a slight deviation in stoichiometry leads to doping. Bringing the Dirac crossings to the Fermi level is an essential aspect of the broader application of this system. The additional electrons, which assist in bringing the Dirac point to the Fermi level,  could be provided by chemical substitution (for example, Sr$_x$La$_{1-x}$NbO$_3$), by gating or by charge transfer due to the proximity effect. Another option to move the crossing points closer to the Fermi level is to drive SrNbO$_3$ to be more correlated. 

\section{Summary}

We studied the effect of octahedral rotations on electronic structure in SrNbO$_3$ films. Our DFT calculations show that two types of octahedral tilting,  $a^0a^0c^+$ and $a^0a^0c^-$,  are possible and degenerated with the non-tilted phase but favored by compressive strain. Both of them result in the emergence of symmetry-protected Dirac crossings near the Fermi level. Through the analysis of ARPES spectra, we identify that SrNbO$_3$ films exhibit in-phase $a^0a^0c^+$ octahedral tilting, which gives rise to symmetry-protected Dirac nodal lines in the band structure. 
Furthermore, the first principles-based calculations show that when time reversal symmetry is broken (with the magnetic field), the electronic structure of $a^0a^0c^+$ phase hosts considerable Berry curvatures. 
Our study demonstrates the potential of engineering octahedral rotation to create and alter the topology of transition metal oxides.

\section{Acknowledgments}

This project has been supported by the European Union’s Horizon 2020 research and innovation program under the Marie Skłodowska-Curie grant agreement No 884104 (PSI-FELLOW-III-3i).N.P. acknowledges the support of Novo Nordisk Foundation Challenge Program 2021: Smart nanomaterials for applications in life-science, BIOMAG Grant NNF21OC0066526, the support from the ERC Advanced “NEXUS” Grant 101054572 and the Danish Council for Independent Research Technology and Production Sciences for the DFF- Research Project 3 (grant No 00069B). W.H.B. acknowledges the financial support from CNPq (in particular Grant 402919/2021-1) and the computational centers: National Laboratory for Scientific Computing (Santos Dumont - LNCC/MCTI, Brazil) and the Brazilian National Center of High Processing Computing (CENAPAD-SP).

\bibliography{manuscript_ref}

\end{document}